\documentclass{elsart}%
\usepackage{amsfonts}
\usepackage{amsmath}
\usepackage{amssymb}
\usepackage{natbib}
\usepackage{graphicx}%
\begin{document}
\begin{frontmatter}
\title{Analytical calculation of the solid angle defined by a cylindrical detector and a point cosine source with parallel axes\thanksref{FCT}}
\thanks[FCT]{Partially supported by Funda\c{c}\~{a}o para a Ci\^{e}ncia e Tecnologia
(Programa Praxis XXI - BD/15808/98)}
\author{M. J. Prata\thanksref{TelFax}}
\thanks[TelFax]{Tel.: +351-21-944-0690; fax: +351-21-846-3276.}
\ead{mjprata@sapo.pt}
\address{Instituto  Tecnol\'{o}gico e Nuclear (ITN), Estrada Nacional 10, Sacav\'{e}m 2686-953, Portugal}
\begin{abstract}
We derive analytical expressions for the solid angle subtended by a
right finite circular cylinder at a point source with cosine angular distribution in the case where the source direction is parallel to the cylinder axis. As a subsidiary result,
an expression for the solid angle subtended by a disc detector at a spread disc source
is also provided, in the case where the two discs have a common symmetry axis which is also coincident with the source direction.
\end{abstract}
\end{frontmatter}

\section{Introduction}

In many situations in radiation physics the value of the solid angle subtended
by a circular cylindrical detector at a point source is needed. The case of an
isotropic point source has been treated to great extent (\citet{Jaff54},
\citet{Mack56}, \citet{Mask56}, \citet{Mask57}, \citet{Gill70},
\citet{Gard71}, \citet{Gree74}, \citet{Prata2003b}).

In a recent work \citep{Prata2003a} we derived analytical expressions for the
solid angle subtended by a cylinder at a point cosine source, under the
restriction that the source and cylinder axes are orthogonal to each other. In
the present work we obtain similar expressions in the case where the source
direction is parallel to the cylinder axis. An expression (eq.
\ref{eq_omega_circ_Hubbell}) for the solid angle subtended by a circular disc
with symmetry axis parallel to the source direction is also given as an
auxiliary result. It should be mentioned that this latter expression appeared
previously \citep[][eq. 29]{Hubb61} in a slightly different context. The
authors detailed a quite general procedure to calculate the response of a
small detector to an axially symmetric source with arbitrary polar angle
distribution, uniformly spread on a circular disc. In that work,
$\Omega_{circ}$ (here eq. \ref{eq_omega_circ_Hubbell}) is interpreted as the
response of a plane detector parallel to a Lambertian uniformly distributed
disc source. In the same work the result is also credited to other authors
(\citet{Herm00}, \citet{Foot15}).

With the on-going computer revolution, very complex problems otherwise
impossible to tackle, are routinely addressed using numerical methods.
Calculations such as the one presented here, amounting to a one-dimensional
integration, can be quickly specified and performed in a desk-top computer
with minimal effort\footnote{In fact, some of the results displayed in the
present work were checked against a numerical integration.}. Nevertheless, we
believe that in the few fortunate occasions where simple expressions exist in
closed form, they deserve an interested look and are worth the effort of
finding them. There are various reasons for this. No matter how careful the
error-checking procedure followed, any numerical result should always be taken
with some degree of prudence. In fact, except for the most simple situations,
a great deal of effort is normally put into testing the result, rather than
into obtaining it. Analytical results can be of some help here by providing
reliable tests to computational algorithms. Also, a closed expression can
sometimes be given an intuitive interpretation like the one provided for
$\Omega_{circ}$ in section (\ref{section_GeoInterp_OmegaCirc}) and, at the
very least, enables the usage of analytical tools to search for extreme
values, regions of monotony, reliable approximations, asymptotic expressions,
etc. Furthermore, the existence of a closed expression for a point source
reduces the complexity of the problem when considering the more realistic case
of a planar spread source. The required four-dimensional integral can of
course be reduced\footnote{In the case of an uniformly spread source on a
disc, a reduction to a double integral could also be achieved with resort to
the exact expressions obtained by \citet{Hubb61} for the radiation field
created by such source at an arbitrary point. The remaining double integral
would then be performed over the detector surface rather than over the source
surface.} to a double integral by using the exact expressions deduced here for
a point source. In section (\ref{section_spreadsource}) we illustrate this by
obtaining an analytical result for the solid angle defined by a cosine source
evenly distributed on a circular disc and a disc detector, in the very special
situation where the two discs share the same symmetry axis which is also
assumed to be coincident with the source direction. Reducing the
dimensionality of a numerical integration is of obvious advantage because the
calculation is less prone to numerical errors and can be performed faster. In
fact, when using quadrature formulae like Simpson's or Gaussian rules to
calculate multidimensional integrals, the total number of points ($N_{T}$)
scales like $N_{T}=N^{d}$, where $N$ is the number of quadrature points used
in each dimension and $d$ is the dimensionality of the integral. The number of
evaluations of the integrand can then become prohibitively high, regardless of
the computing resources available, making necessary the resort to the Monte
Carlo method, which, of course, has the drawback of a small\footnote{It should
be mentioned that there is always some value of $d$ above which the Monte
Carlo method is faster than any quadrature rule. This is so because the
convergence rate of the quadrature rule is that of the one-dimensional rule.
For instance, Simpson's rule error is proportional $N^{-4}$ or to
$N_{T}^{-4/d}$ in $d$ dimensions, which means Simpson's rule should, in
general, be slower than Monte Carlo for $d>8$.} convergence rate (only as
$1/\sqrt{N_{T}}$). This was discussed in a review by \citet{Jame80}, who set
(in 1980) the feasibility limit to the use of a ten-point Gaussian rule to
five dimensions (with limited computer resources available) and to ten
dimensions (with 'unlimited' computer resources).

To illustrate the behavior of the solid angle sample plots are presented in
section (\ref{section_Results}).

\section{Solid Angle Calculation\label{section_solid_angle}}

The solid angle ($\Omega_{surf}$) subtended by a given surface at a point
source can be defined as%

\begin{equation}
\Omega_{surf}=\iint\limits_{\substack{directions\\hitting~surface}%
}f(\mathbf{\Omega})d\Omega~,
\end{equation}
where $f(\mathbf{\Omega})d\Omega$ is the source distribution. In the case of a
point cosine the distribution is defined with respect to some direction axis
specified by the unit vector $\mathbf{k}$ and it is given by $f(\mathbf{\Omega
})=(\mathbf{\Omega\cdot k+}\left\vert \mathbf{\Omega\cdot k}\right\vert
)/(2\pi)$. The $(2\pi)^{-1}$ factor ensures that $0\leq\Omega_{surf}\leq1$.
Setting $\mu=\mathbf{\Omega\cdot k}=\cos(\nu)$, where $\nu$ is the angle
between the two axes, there results that $f(\mathbf{\Omega})=\{\mu/\pi
~(\mu\geq0);0~(\mu<0)\}$ so that the source only emits into the hemisphere
around $\mathbf{k}$. In the following we shall consider the situation of a
right circular cylinder with axis parallel to $\mathbf{k}$. The source is
assumed to be at the origin of the coordinate system and the $z$ axis is
chosen both aligned with $\mathbf{k}$ and parallel to the cylinder axis. The
solid angle is then given by%
\begin{align}
\Omega_{surf} &  =\pi^{-1}{\textstyle\int\nolimits_{\varphi_{\min}}%
^{\varphi_{\max}}}{\textstyle\int\nolimits_{\theta_{\min}}^{\theta_{\max}}%
}\cos(\theta)\sin(\theta)d\theta d\varphi\nonumber\\
&  =(2\pi)^{-1}{\textstyle\int\nolimits_{\varphi_{\min}}^{\varphi_{\max}}%
}(\sin^{2}(\theta_{\max})-\sin^{2}(\theta_{\min}))d\varphi
~,\label{eq_omega_surf}%
\end{align}
where $\theta$ is the polar angle from the $z$ axis; $\varphi$ is the
azimuthal angle in the xy plane and the limiting angles are to be determined
from the conditions that $\mu=\cos(\theta)\geq0$ and that each included
$(\theta,\varphi)$ direction hits the surface.

\begin{figure}[h]
\begin{center}
\includegraphics[
height=4.0154cm,
width=4.3713cm
]{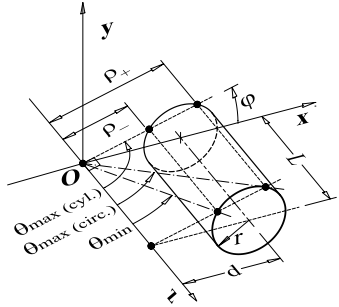}
\end{center}
\caption{Notation for $\Omega_{cyl0}$ and for $\Omega_{circ}$ ($z=L$) when
$d>r$}%
\label{fig1}%
\end{figure}

\begin{figure}[h]
\begin{center}
\includegraphics[
height=4.0154cm,
width=4.0374cm
]{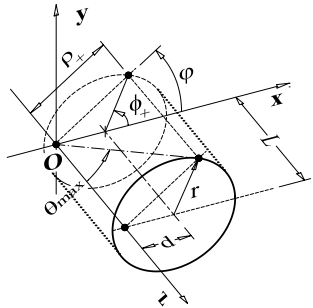}
\end{center}
\caption{Notation for $\Omega_{circ}$ when $d<r$}%
\label{fig2}%
\end{figure}

The solid angle $\Omega$ of the whole cylinder can in general be decomposed
according to $\Omega=\Omega_{cyl}+\Omega_{circ}$, where $\Omega_{cyl}$ and
$\Omega_{circ}$ are the contributions of the cylindrical surface and of one of
the end circles. To calculate these quantities we refer to figs. \ref{fig1}
and \ref{fig2}, where, for simplicity, we assume that the cylinder and disc
axes lie in the $xz$ plane. To obtain $\Omega_{cyl}$ it is sufficient to
consider the situation depicted in fig. \ref{fig1}, where one of the end discs
is at the same plane as the source (i.e. $z=0$). Let $\Omega_{cyl0}(L,r,d)$
denote the solid angle in this case and $\Omega_{circ}(L,r,d)$ the solid angle
defined by the disc. In the case of the disc we distinguish the situation
shown if fig. \ref{fig2}, where $r>d$, from that where $d>r$. For economy this
latter situation is also represented in fig. \ref{fig1}, with reference to the
circle at the plane $z=L$.

\begin{figure}[h]
\begin{center}
\includegraphics[
height=4.5887cm,
width=9.3686cm
]{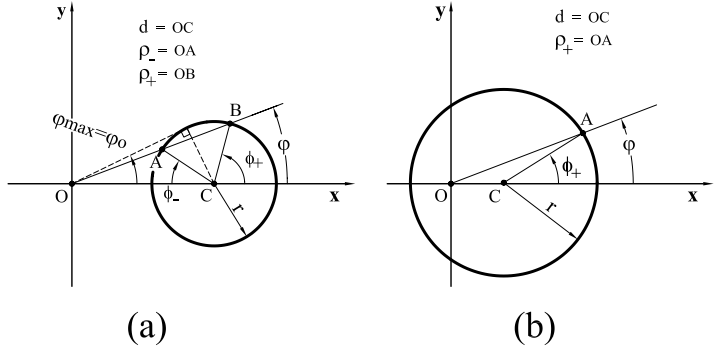}
\end{center}
\caption{Definitions of (a) $\rho_{\pm}$ , $\phi_{\pm}$ and $\varphi_{o}$
($d>r$)\ and (b) $\rho_{+}$ , $\phi_{+}$ ($d<r$)\ }%
\label{fig3}%
\end{figure}

Using eq. \ref{eq_omega_surf} it follows that%

\begin{align}
\Omega_{cyl0}(L,r,d)  &  =\pi^{-1}\int\limits_{0}^{\varphi_{o}}(1-\rho_{-}%
^{2}(\varphi)/[L^{2}+\rho_{-}^{2}(\varphi)])~d\varphi\nonumber\\
&  =\pi^{-1}\int\limits_{0}^{\varphi_{o}}L^{2}/[L^{2}+\rho_{-}^{2}%
(\varphi)]~d\varphi\equiv\pi^{-1}A_{-}(\varphi_{o})
\label{eq_omega_cyl0_a_minus}%
\end{align}

\begin{align}
\Omega_{circ}(L,d  &  >r)=\pi^{-1}\int\limits_{0}^{\varphi_{o}}(\rho_{+}%
^{2}(\varphi)/[L^{2}+\rho_{+}^{2}(\varphi)]-\rho_{-}^{2}(\varphi)/[L^{2}%
+\rho_{-}^{2}(\varphi)])~d\varphi\nonumber\\
&  =\pi^{-1}(\int\limits_{0}^{\varphi_{o}}L^{2}/[L^{2}+\rho_{-}^{2}%
(\varphi)]~d\varphi-\int\limits_{0}^{\varphi_{o}}L^{2}/[L^{2}+\rho_{+}%
^{2}(\varphi)]~d\varphi)\nonumber\\
&  \equiv\pi^{-1}[A_{-}(\varphi_{o})-A_{+}(\varphi_{o})]
\label{eq_omega_circ_a_minus_plus_d_bigger}%
\end{align}

\begin{align}
\Omega_{circ}(L,d  &  <r)=\pi^{-1}\int\limits_{0}^{\pi}\rho_{+}^{2}%
(\varphi)/[L^{2}+\rho_{+}^{2}(\varphi)]~d\varphi\nonumber\\
&  =1-\pi^{-1}\int\limits_{0}^{\pi}L^{2}/[L^{2}+\rho_{+}^{2}(\varphi
)]~d\varphi\nonumber\\
&  =1-\pi^{-1}[A_{+}(\pi)] \label{eq_omega_circ_a_minus_plus_r_bigger}%
\end{align}

where, from fig.\ref{fig3},%

\begin{equation}
\varphi_{o}\equiv\arcsin(r/d)~ \label{eq_phio_def}%
\end{equation}

and%

\begin{equation}
\rho_{\pm}(\varphi)=d\cos\varphi\pm\sqrt{r^{2}-\left(  d\sin\varphi\right)
^{2}}~. \label{eq_rho_plus_minus_def}%
\end{equation}

The required integrals $A_{\pm}$ can be expressed in terms of the integral%

\begin{equation}
I(L,r,d,\phi_{+})=\int\frac{L^{2}}{L^{2}+\rho_{+}^{2}(\phi_{+})}\frac{1}%
{2}(1+\frac{r^{2}-d^{2}}{\rho_{+}^{2}(\phi_{+})})~d\phi_{+}~, \label{eq_I_def}%
\end{equation}

where%

\begin{equation}
\rho_{+}(\phi_{+})=\sqrt{d^{2}+r^{2}+2dr\cos(\phi_{+})}
\label{eq_rho_plus_phi}%
\end{equation}

To proceed we begin by calculating $I$.

\subsection{Calculation of $I$}

\bigskip The integrand in the rhs of eq. \ref{eq_I_def} can be written as%

\begin{equation}
\frac{1}{2}(\frac{L^{2}+d^{2}-r^{2}}{L^{2}+d^{2}+r^{2}}\frac{1}{1+m\cos
(\phi_{+})}+\frac{r^{2}-d^{2}}{d^{2}+r^{2}}\frac{1}{1-n\cos(\phi_{+})})~,
\end{equation}

where%

\begin{equation}
m=2rd/(L^{2}+d^{2}+r^{2})~ \label{eq_m_def}%
\end{equation}

and%

\begin{equation}
n=2rd/(d^{2}+r^{2})~. \label{eq_n_def}%
\end{equation}

The integration is straightforward, giving%

\begin{align}
I  &  =\frac{L^{2}+d^{2}-r^{2}}{L^{2}+d^{2}+r^{2}}\frac{1}{\sqrt{1-m^{2}}%
}\arctan[\sqrt{\frac{1-m}{1+m}}\tan(\frac{\phi_{+}}{2})]\nonumber\\
&  +\frac{r^{2}-d^{2}}{d^{2}+r^{2}}\frac{1}{\sqrt{1-n^{2}}}\arctan[\sqrt
{\frac{1-n}{1+n}}\tan(\frac{\phi_{+}}{2})]~.
\end{align}

Since%

\[
(r^{2}-d^{2})/(d^{2}+r^{2})=\left\{  \sqrt{1-n^{2}}~(r>d);-\sqrt{1-n^{2}%
}~(d>r)\right\}
\]

and%

\begin{gather}
(L^{2}+d^{2}-r^{2})/(L^{2}+d^{2}+r^{2})=\nonumber\\
\left\{  1-m/n(1+\sqrt{1-n^{2}})~(r>d);1-m/n(1-\sqrt{1-n^{2}})~(d>r)\right\}
\end{gather}

there results that%

\begin{align}
I  &  =\frac{1-m/n(1+\sqrt{1-n^{2}})}{\sqrt{1-m^{2}}}\arctan[\sqrt{\frac
{1-m}{1+m}}\tan(\frac{\phi_{+}}{2})]\nonumber\\
&  +\arctan[\sqrt{\frac{1-n}{1+n}}\tan(\frac{\phi_{+}}{2})]~;(r>d)
\label{Eq_I_m_n_r_bigger}%
\end{align}

and%

\begin{align}
I  &  =\frac{1-m/n(1-\sqrt{1-n^{2}})}{\sqrt{1-m^{2}}}\arctan[\sqrt{\frac
{1-m}{1+m}}\tan(\frac{\phi_{+}}{2})]\nonumber\\
&  -\arctan[\sqrt{\frac{1-n}{1+n}}\tan(\frac{\phi_{+}}{2})]~;(d>r)~.
\label{Eq_I_m_n_d_bigger}%
\end{align}

\bigskip

To express $A_{+}$ in the rhs of eqs.
\ref{eq_omega_circ_a_minus_plus_d_bigger} and
\ref{eq_omega_circ_a_minus_plus_r_bigger} in terms of $I$, a change of
variable to $\phi_{+}$ represented in fig. \ref{fig3} is made, using $\phi
_{+}/2=\arctan\left[  \sin(\varphi)\rho_{+}/(r-d+\cos(\varphi)\rho
_{+})\right]  $, where $\rho_{+}=\rho_{+}(\varphi)$ is obtained from eq.
\ref{eq_rho_plus_minus_def}. It follows that $A_{+}=I$ and, changing the
integration limits, that%

\begin{equation}
A_{+}(\pi)=I|_{0}^{\pi} \label{eq_a_plus_I_pi}%
\end{equation}
and%

\begin{equation}
A_{+}(\varphi_{o})=I|_{0}^{\pi/2+\varphi_{o}}~. \label{eq_a_plus_I_phi}%
\end{equation}

In the case of $A_{-}$, the integration variable is first changed to $\phi
_{-}$ shown in fig. \ref{fig3} and given by $\phi_{-}/2=\arctan\left[
\sin(\varphi)\rho_{-}/(r+d-\cos(\varphi)\rho_{-})\right]  $, where, again,
$\rho_{-}$ is defined through eq. \ref{eq_rho_plus_minus_def}. Then,

$A_{-}(\varphi_{o})=-\int\limits_{0}^{\pi/2-\varphi_{o}}\frac{L^{2}}%
{L^{2}+\rho_{-}^{2}(\phi_{-})}\frac{1}{2}(1+\frac{r^{2}-d^{2}}{\rho_{-}%
^{2}(\phi_{-})})~d\phi_{-}~,$

where%

\begin{equation}
\rho_{-}(\phi_{-})=\sqrt{d^{2}+r^{2}-2dr\cos(\phi_{-})}~.
\label{eq_rho_minus_phi}%
\end{equation}

Making a further change to $\widetilde{\phi}=\pi-\phi_{-}$ yields%

\begin{equation}
A_{-}(\varphi_{o})=-\int\limits_{\pi/2+\varphi_{o}}^{\pi}\frac{L^{2}}%
{L^{2}+\rho_{+}^{2}(\widetilde{\phi})}\frac{1}{2}(1+\frac{r^{2}-d^{2}}%
{\rho_{+}^{2}(\widetilde{\phi})})~d\widetilde{\phi}~,
\label{eq_a_minus_phi_tilde}%
\end{equation}

where eqs. \ref{eq_rho_minus_phi} and \ref{eq_rho_plus_phi} were used to write

$\rho_{-}(\phi_{-})=\rho_{-}(\pi-\widetilde{\phi})=\sqrt{d^{2}+r^{2}%
+2dr\cos(\widetilde{\phi})}=\rho_{+}(\widetilde{\phi})~.$

By comparison of eqs. \ref{eq_a_minus_phi_tilde} and \ref{eq_I_def} :%

\begin{equation}
A_{-}(\varphi_{o})=-I|_{\pi/2+\varphi_{o}}^{\pi}~. \label{eq_a_minus_I_phi}%
\end{equation}

Using eqs. \ref{eq_a_minus_I_phi}, \ref{eq_a_plus_I_phi} and
\ref{eq_a_plus_I_pi}, eqs. \ref{eq_omega_cyl0_a_minus} to
\ref{eq_omega_circ_a_minus_plus_r_bigger} can be written as%

\begin{equation}
\Omega_{cyl0}(L,r,d)=-\pi^{-1}I|_{\pi/2+\varphi_{o}}^{\pi}~,
\end{equation}

\begin{equation}
\Omega_{circ}(L,d>r)=-\pi^{-1}(I|_{\pi/2+\varphi_{o}}^{\pi}+I|_{0}%
^{\pi/2+\varphi_{o}})=-\pi^{-1}I|_{0}^{\pi} \label{eq_omega_circ_I_d_bigger}%
\end{equation}

and%

\begin{equation}
\Omega_{circ}(L,r>d)=1-\pi^{-1}I|_{0}^{\pi}~. \label{eq_omega_circ_I_r_bigger}%
\end{equation}

Since $I$ is discontinuous at $d=r$ (see eqs. \ref{Eq_I_m_n_r_bigger},
\ref{Eq_I_m_n_d_bigger}), it should be clear that eqs.
\ref{eq_omega_circ_I_d_bigger} and \ref{eq_omega_circ_I_r_bigger} do
\textit{not} mean that $\Omega_{circ}(L,d>r)=\Omega_{circ}(L,r>d)-1$.

We now turn to eqs. \ref{Eq_I_m_n_d_bigger} and \ref{Eq_I_m_n_r_bigger} to
evaluate the integrals. Since $\tan(\phi_{+}/2)|_{\phi_{+}=\pi/2+\varphi_{o}%
}=[(d+r)/(d-r)]^{1/2}=[(1+n)/(1-n)]^{1/4}$ and making use of $\arctan
(z)+\arctan(1/z)=\pi/2~,(z>0)$, one obtains%

\begin{gather}
\Omega_{cyl0}(L,r,d)=\nonumber\\
\pi^{-1}(\arctan\left[  \sqrt[4]{\frac{1+n}{1-n}}\right]  -\frac
{1-m/n(1-\sqrt{1-n^{2}})}{\sqrt{1-m^{2}}}\arctan\left[  \sqrt{\frac{1+m}{1-m}%
}\sqrt[4]{\frac{1-n}{1+n}}\right]  )~. \label{eq_omega_cylo_m_n}%
\end{gather}

For $\Omega_{circ}$ there results%

\begin{equation}
\Omega_{circ}(L,d>r)=1/2(1-\frac{1-m/n(1-\sqrt{1-n^{2}})}{\sqrt{1-m^{2}}})
\label{eq_omega_circ_m_n_d_bigger}%
\end{equation}

\begin{equation}
\Omega_{circ}(L,d<r)=1/2(1-\frac{1-m/n(1+\sqrt{1-n^{2}})}{\sqrt{1-m^{2}}})
\label{eq_omega_circ_m_n_r_bigger}%
\end{equation}

Defining $\beta$ and $\gamma$ by%

\begin{equation}
\tan(\beta/2)=\sqrt[4]{(1+n)/(1-n)} \label{eq_beta_def}%
\end{equation}

and \bigskip%
\begin{equation}
\cos(\gamma)=m~, \label{eq_gamma_def}%
\end{equation}

eqs. \ref{eq_omega_cylo_m_n}, \ref{eq_omega_circ_m_n_d_bigger} and
\ref{eq_omega_circ_m_n_r_bigger} can be recast as%

\begin{equation}
\Omega_{circ}(L,d>r)=1/2(1-\frac{1+\cos(\beta)\cos(\gamma)}{\sin(\gamma)})~,
\label{eq_omega_circ_beta_gamma_dbigger}%
\end{equation}
\bigskip%
\begin{equation}
\Omega_{circ}(L,r>d)=1/2(1-\frac{\cos(\beta)+\cos(\gamma)}{\sin(\gamma
)\cos(\beta)}) \label{eq_omega_circ_beta_gamma_rbigger}%
\end{equation}

and%

\begin{gather}
\Omega_{cyl0}(L,r,d)=\nonumber\\
\pi^{-1}(\frac{\beta}{2}-\frac{1+\cos(\beta)\cos(\gamma)}{\sin(\gamma)}%
\arctan\left[  \cot(\frac{\gamma}{2})\cot(\frac{\beta}{2})\right]  )~.
\label{eq_omega_cylo_beta_gamma}%
\end{gather}

From eq. \ref{eq_beta_def} it is seen that $\pi/2\leq\beta\leq\pi$ and that,
for $d>r$, $\beta=\pi/2+\varphi_{o}$.

As functions of $L$, $d$ and $r$, eqs. \ref{eq_omega_circ_m_n_d_bigger} and
\ref{eq_omega_circ_m_n_r_bigger} can be rewritten to give $\Omega_{circ}$ in
terms of a single expression:
\begin{equation}
\Omega_{circ}(L,d,r)=1/2(1-\frac{d^{2}+L^{2}-r^{2}}{\sqrt{(r^{2}+d^{2}%
+L^{2})^{2}-4r^{2}d^{2}}})~, \label{eq_omega_circ_Hubbell}%
\end{equation}
which is equivalent to \citet[][eq. 29]{Hubb61}, except for a factor of $\pi$.

\subsection{Special values and continuity}

First, the cases $L=0$ and $d=r$ are discussed. From eqs. \ref{eq_m_def} and
\ref{eq_n_def}, it follows that

$L\rightarrow0\Leftrightarrow m\rightarrow n~,$

$d\rightarrow r\Leftrightarrow\left\{  n\rightarrow1\wedge m\rightarrow
m_{1}\right\}  ~,$

where $m_{1}=2r^{2}/(L^{2}+2r^{2})~.$

Then, using eq. \ref{eq_omega_cylo_m_n},

$\Omega_{cyl0}(L\rightarrow0,r<d)=\Omega_{cyl0}(m\rightarrow n)=0$

$\Omega_{cyl0}(L\neq0,d\rightarrow r^{+})=\Omega_{cyl0}(n\rightarrow
1,m\rightarrow m_{1})=1/2$

and $\Omega_{cyl0}$ is thus discontinuous when $L\rightarrow0$, $d\rightarrow
r^{+}$.

Starting from eqs. \ref{eq_omega_circ_m_n_d_bigger},
\ref{eq_omega_circ_m_n_r_bigger},

$\Omega_{circ}(L\neq0,d\rightarrow r^{\pm})=\Omega_{circ}(n\rightarrow
1,m\rightarrow m_{1})=1/2(1-\sqrt{\frac{1-m_{1}}{1+m_{1}}})$

$\Omega_{circ}(L\rightarrow0)=\left\{  0~(d>r),1/2~(d=r),1~(d<r)\right\}  $.
One concludes that $\Omega_{circ}$ is continuous whenever $L\neq0$.

Finally, for an infinite length cylinder at a skew position from the source
(i.e. $d>r$) and with one end at the source plane there is the upper
asymptotic value%

\begin{gather}
\Omega_{cyl0}(L\rightarrow\infty,r<d)=\Omega_{cyl0}(m\rightarrow
0)=1/2-2/\pi\arctan(\sqrt[4]{\frac{1-n}{1+n}}%
)\label{eq_omega_cylo_assimptotic}\\
=\varphi_{o}/\pi~. \label{eq_omega_cylo_assimptotic_phi0}%
\end{gather}

The last result can be checked directly using eq. \ref{eq_omega_cyl0_a_minus}
and can also be interpreted as the solid angle defined by an infinite length rectangle.

\subsection{Geometrical interpretation of $\ \Omega_{circ}%
\label{section_GeoInterp_OmegaCirc}$}

\begin{figure}[h]
\begin{center}
\includegraphics[
height=3.7606cm,
width=4.0857cm
]{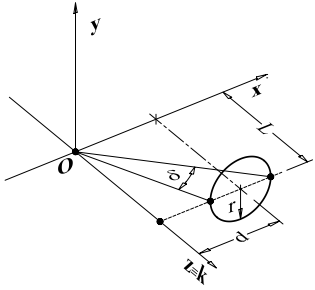}
\end{center}
\caption{Definition of $\delta$.}%
\label{fig4}%
\end{figure}

$\Omega_{circ}$ can be given a simple geometrical meaning by noticing that%
\begin{equation}
\Omega_{circ}=1/2(1-\cos(\delta))~,\label{eq_omega_circ_delta}%
\end{equation}
where $\delta$, shown in fig. \ref{fig4}, is the the aperture of the solid
angle cone measured in the plane containing the source axis and the disc
center. Eq. \ref{eq_omega_circ_delta} is easily proved by writing
$\delta=\arctan[(d+r)/L]-\arctan[(d-r)/L]=\arctan[2rL/(L^{2}+d^{2}-r^{2})]$,
where in the last step we used $arctan(z_{1})-arctan(z_{2})=arctan[(z_{1}%
-z_{2})/(1+z_{1}z_{2})]$. There results that $\cos(\delta)=(L^{2}+d^{2}%
-r^{2})/\sqrt{(L^{2}+d^{2}-r^{2})^{2}+(2Lr)^{2}}$. Since $(L^{2}+d^{2}%
-r^{2})^{2}+(2Lr)^{2}=(L^{2}+d^{2}+r^{2})^{2}-(2dr)^{2}$, it is seen that eq.
\ref{eq_omega_circ_delta} follows directly from eq.
\ref{eq_omega_circ_Hubbell}.

\subsection{The solid angle for a parallel coaxial disc
source\label{section_spreadsource}}

\begin{figure}[h]
\begin{center}
\includegraphics[
height=4.2241cm,
width=4.0857cm
]{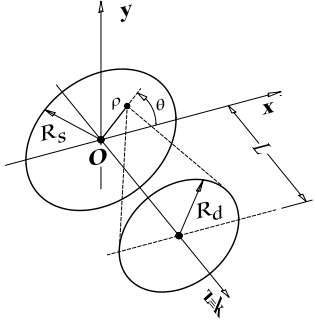}
\end{center}
\caption{Notation for the coaxial , parallel, circular source and detector
set.}%
\label{fig5}%
\end{figure}

The solid angle defined by a spread cosine source evenly distributed on a disc
and a coaxial, parallel, disc detector can also be obtained in closed form.
Referring to fig. \ref{fig5}, we see that the solid angle is given by%

\begin{equation}
\Omega_{s}=\frac{1}{\pi R_{s}^{2}}\int_{0}^{R_{s}}\int_{0}^{2\pi}\rho~
\Omega_{circ}(L,R_{d},\rho)d\theta d\rho~,
\end{equation}
where $R_{s}$ and $R_{d}$ are the source and detector radii; and
$\Omega_{circ}(L,R_{d},\rho)\rho d\theta d\rho$ is the contribution of an
elemental area which can be obtained by putting $r=R_{d}$ and $d=\rho$ in eq.
\ref{eq_omega_circ_Hubbell}. Then,%

\begin{equation}
\Omega_{s}=\frac{1}{2R_{s}^{2}}\left[  \rho^{2}-\sqrt{(L^{2}+\rho^{2}%
-R_{d}^{2})^{2}+(2LR_{d})^{2}}\right]  _{\rho=0}^{\rho=R_{s}}~.
\end{equation}
A little algebra yields%

\begin{equation}
\Omega_{s}(L,R_{d},R_{s})=\frac{R_{d}}{R_{s}}\frac{1}{2m_{s}}\left[
1-\sqrt{1-m_{s}^{2}}\right]  ~, \label{eq_omegaS2}%
\end{equation}
or%
\begin{equation}
\Omega_{s}(L,R_{d},R_{s})=\frac{R_{d}}{R_{s}}\frac{\tan(\gamma_{s}/2)}{2}~,
\end{equation}
where%

\begin{equation}
\sin(\gamma_{s})\equiv m_{s}\equiv\frac{2R_{d}R_{s}}{L^{2}+R_{d}^{2}+R_{s}%
^{2}}~. \label{eq_ms_def}%
\end{equation}

\section{Results and discussion\label{section_Results}}

In order to define the general position of the cylinder, let $L_{1}$, $L_{2}$
be the z coordinates of the end discs and set $L_{1}=L_{2}+length>L_{2}$.

The solid angle $\Omega(L_{1},L_{2},r,d)$ can be calculated using eqs.
\ref{eq_omega_cylo_m_n}, \ref{eq_omega_circ_m_n_d_bigger}
\ref{eq_omega_circ_m_n_r_bigger}, \ref{eq_m_def} and \ref{eq_n_def}. The
several situations to be considered are summarized in\textbf{\ }table
\ref{tab1}.

\begin{table}[h]
\caption{Expressions for the solid angle of the whole detector. $L_{1}$,
$L_{2}$ ($L_{1}>L_{2})$ are the z coordinates of the end discs.}%
\label{tab1}%
\begin{tabular}
[c]{ccc}\hline
$L_{1},L_{2}$ & $d,r$ & $\Omega$\\\hline
$L_{1}<0$ & $-$ & $0^{(a)}$\\
$L_{1}>0,L_{2}<0$ & $d<r$ & $1^{(b)}$\\
& $d>r$ & $\Omega_{cyl0}(L_{1},r,d)$\\
$L_{1}>0,L_{2}>0$ & $d<r$ & $\Omega_{circ}(L_{2},r,d)$\\
& $d>r$ & $\qquad\Omega_{cyl0}(L_{1},r,d)-\Omega_{cyl0}(L_{2},r,d)+\Omega
_{circ}(L_{2},r,d)$\\\hline
\multicolumn{3}{l}{\qquad\vspace{-0.1cm}(a) the source only emits into
$z\geq0$}\\
\multicolumn{3}{l}{\qquad(b) source inside the detector}%
\end{tabular}
\end{table}

As examples, we consider two detectors of radius $1$ and lengths $5$
($L_{1}=L_{2}+5$) and $10$ ($L_{1}=L_{2}+10$).

\begin{figure}[h]
\begin{center}
\includegraphics[
height=4.0857cm,
width=6.3702cm
]{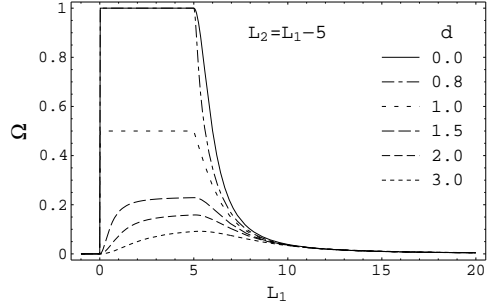}
\end{center}
\caption{Solid angle defined by a cylinder of radius $1$ and length $5$.}%
\label{fig6}%
\end{figure}

\begin{figure}[h]
\begin{center}
\includegraphics[
height=4.0857cm,
width=6.3702cm
]{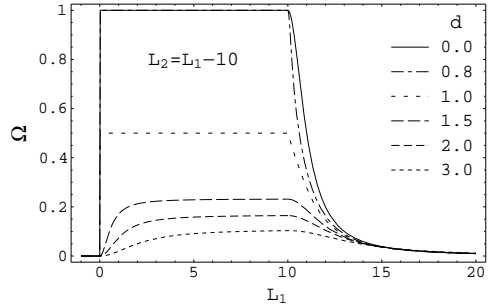}
\end{center}
\caption{Solid angle defined by a cylinder of radius $1$ and length $10$.}%
\label{fig7}%
\end{figure}

Plots of $\Omega$ for each detector as a function of $L_{1}$ are shown in
figs. \ref{fig6} and \ref{fig7}, for different values of distance $d$. When
$d<r$ the curves for the two detectors are essentially the same except for a
displacement equal to the difference of the detectors lengths. This is so
because for $L_{1}$ smaller than the length of the detector ($L_{2}<0$), the
source is inside the detector and $\Omega=1$; for $L_{2}>0$ the solid angle is
that of the circle ($\Omega=\Omega_{circ}$) and it has the same value for both
equal-radius detectors.

When $d>r$, $\Omega$ increases with $L_{1\text{ }}$as the detector is drawn
from 'behind' the source. For the smaller distances (e.g. $d=1.5$) $\Omega$
quickly rises to the asymptotic value (see eq. \ref{eq_omega_cylo_assimptotic}%
) and remains almost constant until $L_{1}$ is increased to values bigger than
the length of the detector ($L_{2}>0$). Then $\Omega$ falls off as the
detector is moved away from the source. It should be clear that $\Omega(d>r)$
is the same for both detectors if $L_{1}$ is smaller than the length of the
shorter detector (i.e. $L_{1}\leq5$).

\begin{ack}
Thanks are due to Jo\~{a}o Prata for reviewing this manuscript.
I would like to thank Professor John H. Hubbell for
providing a copy of the works by A.V. Masket \citep{Mask57}, A.H. Jaffey \citep{Jaff54}
and Hubbell \textit{et al} \citep{Hubb61}.
This work was partially supported by Funda\c{c}\~{a}o para a Ci\^{e}ncia e Tecnologia
(Grant BD/15808/98 - Programa Praxis XXI). Thanks are due to the Referee of this manuscript for
valuable suggestions.
\end{ack}


\begin{thebibliography}{99999999999999999999999999999}                                                                    %


\bibitem[Foote, 1915]{Foot15}Foote, P.D., 1915. Bull. of NBS 12, 583. Cited in \citet{Hubb61}.

\bibitem[Gardner and Verghese, 1971]{Gard71}Gardner, R.P. \& Verghese, K.,
1971. On the solid angle subtended by a circular disc. Nucl. Instr. Meth. 93, 163-167.

\bibitem[Gillespie, 1970]{Gill70}Gillespie, C.R., 1970. Determination of the
geometrical factor of cylindrical geometries. Rev. Sci. Instr. 41 (1), 42-43.

\bibitem[Green \textit{et al}, 1974]{Gree74}Green, M.V., Aamodt, R.L. \&
Johnston, G.S., 1974. The solid angle subtended by a solid, right, circular
cylinder as seen from a point in space. Nucl. Instr. Meth. 117, 409-412.

\bibitem[Herman, 1900]{Herm00}Hermann, R.A.,1900. A treatise in geometrical
optics. Cambridge Univ. Press, 217 (ex. 13). Cited in \citet{Hubb61}.

\bibitem[Hubbell \textit{et al}, 1961]{Hubb61}Hubbell, J.H., Bach, R.L. \&
Herbold, R.J., 1961. Radiation field from a circular disc source. J. Research
NBS 65C (4), 249-264.

\bibitem[Jaffey, 1954]{Jaff54}Jaffey, A.H., 1954. Solid angle subtended by a
circular aperture at point and spread sources: formulas and some tables. Rev.
Sci. Instr. 25 (4), 349-354.

\bibitem[James, 1980]{Jame80}James, F., 1980. Monte Carlo theory and practice.
Rep. Prog. Phys. 43, 1145-1189.

\bibitem[Macklin, 1957]{Mack56}Macklin, P.A., 1957. \textit{Expression for the
solid angle subtended by a circular disc at a point source in terms of
elliptic integrals. Included as a footnote in } \citet{Mask57}.

\bibitem[Masket \textit{et al}, 1956]{Mask56}Masket, A.V., Macklin, R.L. \&
Schmitt, H.W., 1956. Tables of solid angle values and activations. ORNL-2170
(Oak Ridge Nat. Lab., Oak Ridge, Tenn.)

\bibitem[Masket, 1957]{Mask57}Masket, A.V., 1957. Solid Angle contour
integrals, series, and tables. Rev. Sci. Instr. 28 (3), 191-197.

\bibitem[Prata, 2003a]{Prata2003a}Prata, M.J., 2003a. Analytical calculation
of the solid angle defined by a cylindrical detector and a point cosine source
with orthogonal axes. Rad. Phys. Chem. 66 (6), 387-395.e-print: math-ph/0209065.


\bibitem[Prata, 2003b]{Prata2003b}Prata, M.J., 2003b. Solid angle subtended by
a cylindrical detector at a point source in terms of elliptic integrals. Rad.
Phys. Chem. 67 (5), 599-603. e-print: math-ph/0211061.


\bibitem[Verghese \textit{et al}, 1972]{Verg72}Verghese, K., Gardner, R.P., \&
Felder, R.M., 1972. Solid angle subtended by a circular cylinder. Nucl. Instr.
Meth. 101, 391-393.
\end{thebibliography}
\end{document}